\documentclass[manuscript]{emulateapj}
\usepackage{natbib}
\usepackage{graphicx} 
\usepackage{txfonts}
\usepackage[dvips]{color}

\makeatletter

\newcommand{\msun}{{$M_{\odot}$}}

\slugcomment{Accepted for publication in ApJ Letters}
\shorttitle{The stellar velocity dispersion of a compact massive galaxy
at z=1.80} \shortauthors{van de Sande et al.}

\begin{document}

\title{The stellar velocity dispersion of a compact massive galaxy at
$z=1.80$ using X-Shooter: confirmation of the evolution in the 
mass-size and mass-dispersion relations \altaffilmark{1,}\altaffilmark{2}}

\author{Jesse van de Sande\altaffilmark{3}, Mariska
Kriek\altaffilmark{4}, Marijn Franx\altaffilmark{3}, Pieter G. van
Dokkum\altaffilmark{5}, Rachel Bezanson\altaffilmark{5}, Katherine
E. Whitaker\altaffilmark{5}, Gabriel Brammer\altaffilmark{6}, Ivo
Labb\'e\altaffilmark{3}, Paul J. Groot\altaffilmark{7}, Lex
Kaper\altaffilmark{8} }

\altaffiltext{1}{Based on X-Shooter-VLT observations collected at the
European Southern Observatory, Paranal, Chile.}

\altaffiltext{2} {Based on observations with the NASA/ESA
\textit{Hubble Space Telescope (HST)}, obtained at the Space Telescope
Science Institute, which is operated by AURA, Inc., under NASA
contract NAS 5-26555.}

\altaffiltext{3}{Leiden Observatory, Leiden University, P.O.\ Box
9513, 2300 RA Leiden, The Netherlands.}

\altaffiltext{4}{Harvard-Smithsonian Center for Astrophysics, 60
Garden Street, Cambridge, MA 02138, USA}

\altaffiltext{5}{Department of Astronomy, Yale University, P.O. \ Box
208101, New Haven, CT 06520-8101.}

\altaffiltext{6}{European Southern Observatory, Alonso de C\'ordova
3107, Casilla 19001, Vitacura, Santiago, Chile}

\altaffiltext{7}{Department of Astrophysics, IMAPP, Radboud University
Nijmegen, P.O. Box 9010, 6500 GL Nijmegen, The Netherlands}

\altaffiltext{8}{Astronomical Institute Anton Pannekoek, University of
Amsterdam, Science Park 904, 1098 XH Amsterdam, The Netherlands}


\begin{abstract}

Recent photometric studies have shown that early-type galaxies at
fixed stellar mass were smaller and denser at earlier times. In this
paper we assess that finding by deriving the dynamical mass of such a
compact quiescent galaxy at z=1.8. We have obtained a high-quality
spectrum with full UV-NIR wavelength coverage of galaxy NMBS-C7447 using
X-Shooter on the VLT. We determined a velocity dispersion of
294$\pm$51\ $\rm{km~s}^{-1}$. Given this velocity dispersion and the
effective radius of 1.64$\pm$0.15\ kpc (as determined from HST-WFC3
F160W observations) we derive a dynamical mass of 1.7$\pm0.5 \ \times\
$10$^{11}$\msun. Comparison of the full spectrum with stellar
population synthesis models indicates that NMBS-C774 has a relatively
young stellar population ($0.40$ Gyr) with little or no star formation
and a stellar mass of $M_{\star}\sim1.5 \times 10^{11}$\msun.  The
dynamical and photometric stellar mass are in good agreement. Thus,
our study supports the conclusion that the mass densities of quiescent
galaxies were indeed higher at earlier times, and this earlier result
is not caused by systematic measurement errors. By combining available
spectroscopic measurements at different redshifts, we find that the
velocity dispersion at fixed dynamical mass was a factor of $\sim$1.8
higher at z=1.8 compared to z=0.  Finally, we show that the apparent
discrepancies between the few available velocity dispersion
measurements at $z>1.5$ are consistent with the intrinsic scatter of
the mass-size relation.
\end{abstract}


%
\section{Introduction}
\label{sec:introduction}

In hierarchical structure formation models, the most massive
early-type galaxies are assembled last
\citep[e.g.,][]{springel2005}. This simple picture seems difficult to
reconcile with recent studies showing that the first massive, quiescent
galaxies were already in place when the universe was only $\sim$3 Gyr
old (e.g., \citealt{labbe2005}; \citealt{kriek2006};
\citealt{williams2009}). The recent discovery that these high-redshift galaxies still grow significantly in size (e.g.,
\citealt{daddi2005}; \citealt{trujillo2006}; \citealt{vandokkum2008}),
and mass \citep{vandokkum2010} solves this apparent conflict. The
observed compact high-redshift galaxies may simply be the cores of
local massive early-type galaxies, which grow inside-out by accreting
(smaller) galaxies (e.g., \citealt{naab2009}; \citealt{bezanson2009};
\citealt{vanderwel2009}), and thus assemble a significant part of
their mass at later times (see also \citealt{oser2010}).

However, the results may be interpreted incorrectly due to
systematic uncertainties. Firstly, sizes may have been underestimated,
as low-surface brightness components might have been missed
\citep{mancini2010}. Nonetheless, recent work using stacking techniques
(e.g., \citealt{vanderwel2008}; \citealt{cassata2010};
\citealt{vandokkum2010}), and ultra-deep HST-WFC3 data
\citep[e.g.,][]{szomoru2010}, demonstrated that radial profiles can now
be measured with high accuracy extending to large radii.  Secondly,
the stellar mass estimates suffer from uncertainties in stellar
population synthesis (SPS) models, the paucity of spectroscopic
redshifts, and furthermore rely on assumptions regarding the initial
mass function (IMF) and metallicity \citep[e.g.,][]{conroy2009}. Direct
kinematic mass measurements, which are not affected by these
uncertainties, are needed to confirm the high stellar masses and
densities of these galaxies.

Kinematic measurements have only recently become possible for
high-redshift galaxies \citep[e.g.,][]{cenarro2009}. Using optical spectroscopy, \citet{newman2010}
have explored the epoch up to $z\sim1.5$. With near-infrared (NIR)
spectroscopy these studies have been pushed to even higher redshift. Using a $\sim$29\,hr spectrum of an ultra-compact
galaxy at $z=2.2$ obtained with Gemini Near-IR
Spectrograph \citep{kriek2009a},
\citet{vandokkum2009b} found a high, though
uncertain velocity dispersion of $\sigma = 510^{+165}_{-95}$km$~$s$^{-1}$. \citet{onodera2010} used the MOIRCS on the Subaru telescope to observe
the rest-frame optical spectrum of a less-compact, passive,
ultra-massive galaxy at $z=1.82$, but the low spectral resolution only
allowed the determination of an upper limit to the velocity dispersion
of $\sigma < 326$ km s$^{-1}$. With the lack of high-quality dynamical
data at $z>1.5$ there still is no general consensus on the matter of
compact quiescent galaxies.

\begin{figure*}[tbh] \epsscale{1.1} \plotone{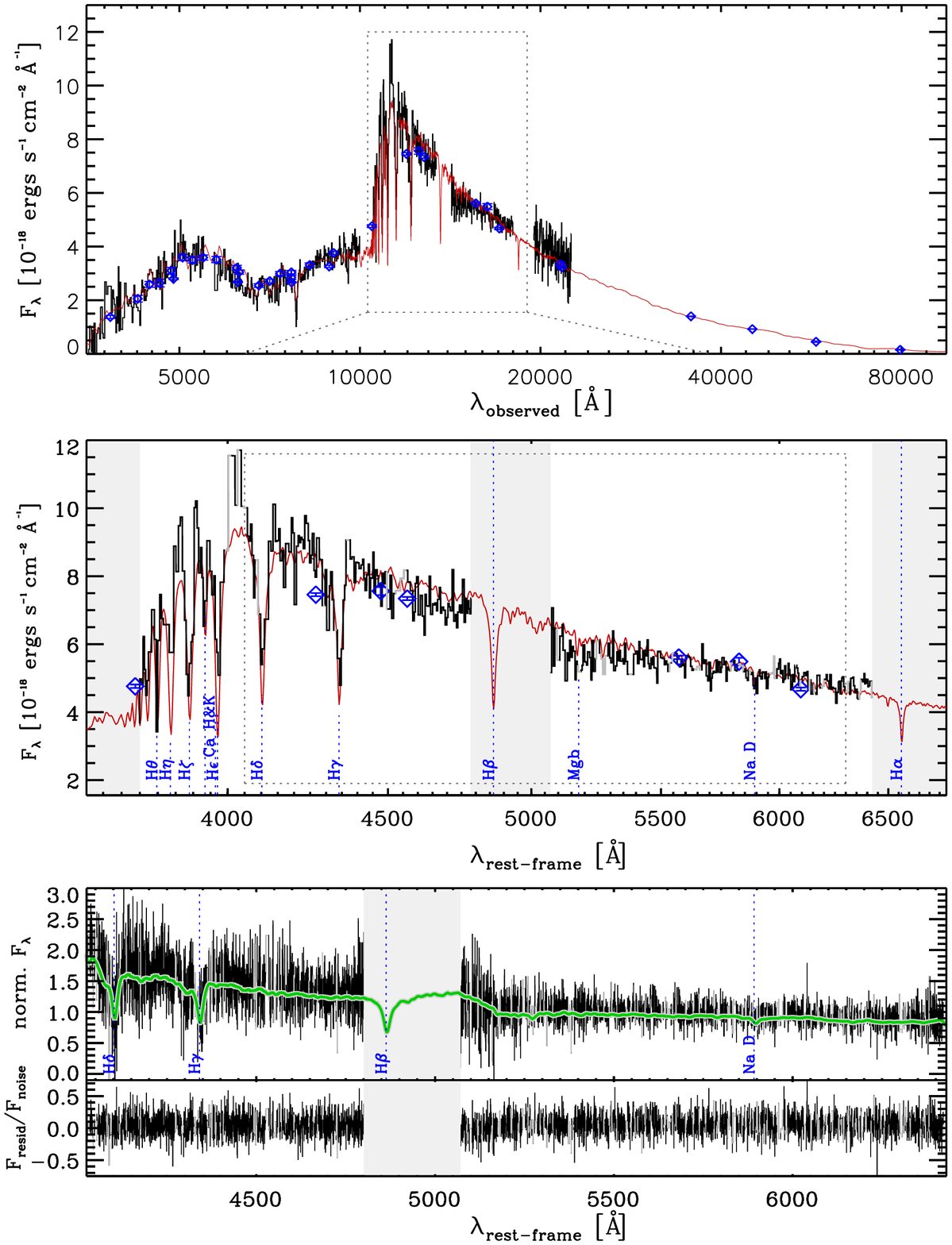}
\caption{X-Shooter spectrum of NMBS-C7447 and the best-fit stellar
population model (red line). \textit{ Top panel:} broad and
medium-band data (blue diamonds) in combination with low-resolution
spectrum (10 $\AA$ per bin). The entire wavelength range from UV
(0.35$\mu$m) to NIR(2.3$\mu$m) is covered in 2 hr integration time
with unprecedented quality. The galaxy is best fit with a young
stellar population ($0.40$ Gyr, $\tau=0.03$Gyr) with little star
formation ($0.002$ \msun yr$^{-1}$), and a stellar mass of
$M_{\star}\sim1.5 \times 10^{11}$ \msun.  \textit{Middle panel:} Zoom
in on the rest-frame optical part of the spectrum. Grey areas indicate
regions of strong skylines or atmospheric absorption. Most prominent
stellar absorption features are indicated with blue dashed
line. \textit{Bottom two panels}: High resolution spectrum (0.5 $\AA$
bin) of the observed features used to determine the stellar velocity
dispersion.  Green line is the best fit for the velocity dispersion
with $4020\ \AA < \lambda_{\rm rest-frame} < 6400\ \AA$ using the pPFX
code \citep{cappellari2004}. The resulting stellar velocity dispersion
is $\sigma_*$= 294$\pm$51 km s$^{-1}$. The residual from the best fit
divided by the noise is shown in the bottom panel.}
\label{fig:fast}
\end{figure*}

Here we present the first high signal-to-noise ratio (S/N),
high-resolution, UV-NIR spectrum of a $z=1.80$ galaxy observed with
X-Shooter \citep{dodorico2006} on the VLT. Throughout the paper we
assume a $\Lambda$CDM cosmology with $\Omega_\mathrm{m}$=0.3,
$\Omega_{\Lambda}=0.7$, and $H_{0}=70$ km s$^{-1}$ Mpc$^{-1}$. All
broadband data are given in the AB-based photometric system.

\section{Observations and Reduction}
\label{sec:observations} The target is selected from the NEWFIRM
Medium-Band Survey (NMBS; \citealt{vandokkum2009a}; \citealt{whitaker2011}).  This target, NMBS-C7447 ($\rm{\alpha=10h 00m
06.955s, \delta=02d 17m 33.603s}$), was selected as it is among the
brightest ($K_{tot} = 19.64$), quiescent galaxies in the COSMOS
field. As the galaxy was selected for its apparent
magnitude, it is probably younger than the typical quiescent galaxy at
its redshift \citep{whitaker2010}. A radio counterpart was detected, with $L=9.789\times10^{24}$
Wm$^{-2}$Hz$^{-1}$ \citep{schinnerer2010}.

The galaxy was observed for 2 hr with
X-Shooter on the VLT/UT2 on January 21st 2010, with clear sky conditions and an
average seeing of 0.8$\arcsec$. X-Shooter
consists of 3 arms: UVB, VIS, and NIR, resulting in a simultaneous
wavelength coverage from 3000 to 24800 $\AA$. The NIR part of the
spectrum is the most interesting, as it covers many of he strong
rest-frame optical stellar absorption features.
A 0.9'' slit was used ($R=5600$ at 1.5 $\mu$m). The 2 hours of observing time were split in 8 exposures of 15
minutes each with an ABA'B' on-source dither pattern. A telluric standard of type B9V was observed for calibration purposes.

We use a similar procedure to reduce cross-dispersed NIR spectra as in
\citet{kriek2008}; details will be given in J. van de Sande et al.\
(in preparation). The resulting 2D spectrum was visually inspected for
emission lines, but none were found.  A 1D spectrum was extracted by
adding all lines (along wavelength direction), with flux greater than
0.1 times the flux in the central row, using optimal weighting. Our results do not change if we take a different flux limit for extraction. This
high-resolution spectrum has a S/N 10.4 $\AA$ in rest frame in $H$.
A low-resolution spectrum was constructed by binning the 2D spectrum
in wavelength direction. Using a bi-weight mean, 20 good pixels,
i.e. not affected by skylines or strong atmospheric absorption, were
combined. The 1D spectrum was extracted from this binned 2D spectrum
in a similar fashion as the high-resolution spectrum (see Figure
\ref{fig:fast}).

For the UVB and VIS arm, the 2D spectra were reduced using the ESO
pipeline (1.2.2, \citealt {goldoni2006}). Correction for the
atmospheric absorption and 1D extraction were performed in a similar
way as the NIR arm as described above.

\begin{figure}[tbh] \epsscale{1.1} \plotone{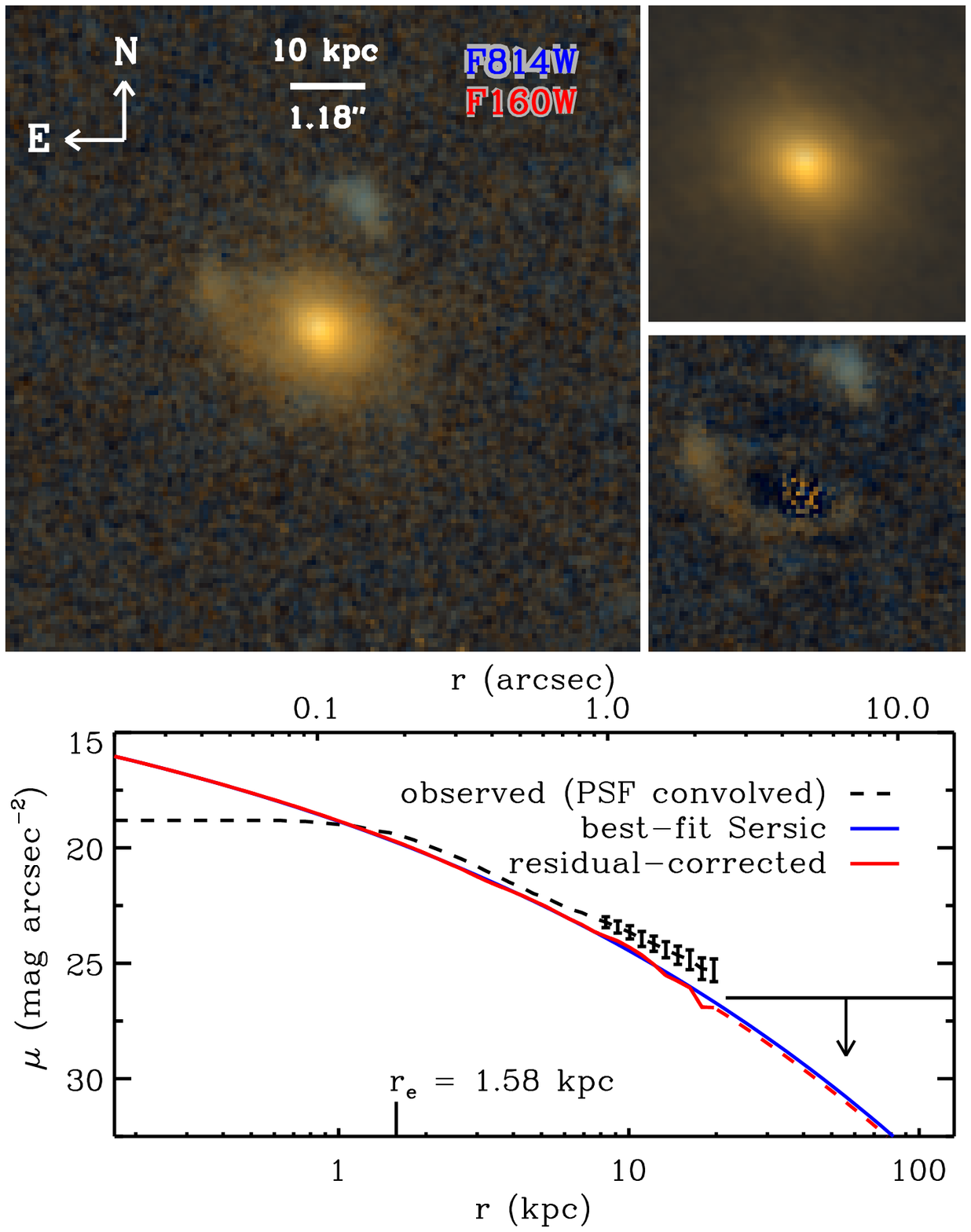}
\caption{\textit{Left panel: }HST-Color image of NMBS-C7296,
consisting of ACS-F775W (blue) and WFC3-F160W (red). \textit{Top
right panel:} best-fit S\'ersic model convolved with the PSF from
GALFIT. The best-fit effective radius is $r_{\rm e}=1.64 \pm 0.15$
kpc and $n=5.3 \pm0.4$ \textit{Bottom right panel: }remaining
residual after subtracting the model from the observed image divided
by the noise.  \textit{Bottom panel}: Observed radial profile of
NMBS-C7447, in comparison with the best-fit S\'ersic profile and the
residual-corrected profile, as described by \citet{szomoru2010}.} 
\label{fig:image}
\end{figure}

\section{Stellar population and Structural Properties }
\subsection{Stellar population properties}
\label{subsec:ssp}

We estimate the redshift, age, dust content, star formation timescale,
metallicity, stellar mass, star formation rate (SFR), and their
confidence intervals by fitting the low-resolution spectrum in
combination with the broadband photometry with SPS models. We use the
fitting code FAST \citep{kriek2009b} in combination with the stellar
templates by \citet{bruzual2003} (BC03) \citep[see][]{kriek2010}. An
exponentially declining star formation history with timescale $\tau$
is assumed, together with a \citet{chabrier2003} IMF, and the
\citet{calzetti2000} reddening law.  We scale the mass by the ratio
between the total F160W flux in the best-fit GALFIT model (see
Section \ref{subsec:sizes}), and the total $H$ band flux in the NMBS
catalog.  The galaxy spectrum is best-fit with a stellar mass of
$M_{\star}=1.5 \times 10^{11}$ \msun, $\tau=0.03$\ Gyr, an age of
$0.40$ Gyr, SFR of $0.002$ \msun yr$^{-1}$, $A_{V}=0.20$, solar metallicity, and a redshift of 1.800 (see Figure \ref{fig:fast}).
In order to account for systematic uncertainties
\citep[e.g.,][]{conroy2009}, we will assume an error of $\sim$0.2 dex
in $M_{\star}$.  The galaxy is not detected at 24 $\mu$m, leading to a
3-$\sigma$ ($\sim$20\ $\mu$Jy) upper-limit to the dust-enshrouded SFR
of $<$\ 15\msun\ yr$^{-1}$.

\subsection{Size measurement}
\label{subsec:sizes}

We obtained HST-WFC3 F160W imaging of NMBS-C7447 in October 2010 (HST-GO-12167.1, see Figure \ref{fig:image}) to measure its size by fitting a S\'ersic
radial surface brightness profile \citep{sersic1968}, using the 2D
fitting program GALFIT (version 3.0.2; \citealt{peng2010}). The blue
object to the North was masked in the fit, as it is unclear whether it
is part of the galaxy. All parameters, including
the sky, were left free for GALFIT to determine, and three nearby
field stars were used for the PSF convolution.

In WFC3 F160W we find a mean circularized effective radius of
1.64$\pm$0.15 kpc, a mean S\'ersic \mbox{$n$-parameter} of
5.3$\pm$0.4, and an axis ratio $b/a=0.71 \pm$0.01. The uncertainties
reflect both sky noise and PSF uncertainties, which were simulated
using different field stars. We find the same effective radius if we
use the residual-corrected method as described by
\citet{szomoru2010}. We also analyzed an ACS I-band image from the
COSMOS survey (\citealt{scoville2007}; \citealt{koekemoer2007}).  The
target has an effective radius of $r_e$=1.95$\pm$0.20 kpc with
$n$=5.6$\pm$0.4, using the same PSF-stars as for WFC3. 

An arclike
feature is present in the residual image, to the South-East of the object
(within 1.5", and $\sim$3 magnitudes fainter than the main
target). This may indicate that the galaxy is undergoing a tidal
interaction (see also \citealt{vandokkum2010b}).

In what follows, we will use the mean effective radius obtained with
WFC3 F160W(\textit{H}), as this band coincides with rest-frame
optical for our $z\sim$1.8 galaxy.

\subsection{Velocity dispersion}
\label{subsec:dispersions} 

We use our high-resolution spectrum, and the Penalized Pixel-Fitting method (pPXF) developed by
\citet{cappellari2004} to measure an accurate stellar velocity dispersion for NMBS-C7447.
Four different templates were used: the best-fit BC03 SPS model
($\sigma$=85 kms$^{-1}$), Munari synthetic stellar library
(\citealt{munari2005}, $\sigma$=6.4 kms$^{-1}$), Indo-US Library
(\citealt{valdes2004}, $\sigma$=38.2 kms$^{-1}$), and the Miles
library (\citealt{miles2006}, $\sigma$=71.9 kms$^{-1}$).  Except for
the best-fit SPS model, pPXF was used to construct an optimal template
in combination with a 30th-order Legendre Polynomial.  The fit was
restricted to $4020 \AA < \lambda <6400 \AA$, in order to exclude the
Balmer break region and the noisier K-band.  Figure \ref{fig:fast}
(bottom panels) shows the high-resolution spectrum with the best-fit
velocity dispersion model from pPXF in red using the best-fit SPS
model.

After correcting for instrumental resolution ($\sigma$=23 km s$^{-1}$)
and the spectral resolution of the templates, we find a best-fitting
velocity dispersion of \mbox{$\sigma_{\rm{obs}}$=284$\pm$51 km
s$^{-1}$}.  The error was determined in the following way.  We
subtracted the best-fit template from the spectrum. This residual was
randomly rearranged in wavelength space and added to best-fit
template.  We determined the velocity dispersion of 1000 simulated
spectra. Our quoted error is the standard deviation of the resulting
distribution of $\sigma$.  When we include the Balmer break region in
the fit, the formal error decreases, but the derived dispersion
becomes very dependent on the chosen stellar template. Fitting the full-wavelength range gives a consistent result of $\sigma_*$=
328$\pm$ 35 km, but we prefer to use the method above as it is the
most robust.

The stellar velocity dispersion is corrected to match the average
dispersion as would be observed within an aperture radius of
$r_e$. Our approach is similar to \citet{cappellari2006}, but taken
into account the effects of a non-circular aperture, seeing, and
optimized extraction. The aperture correction is only 3.5\%, resulting
in a velocity dispersion of \mbox{$\sigma_e$=294$\pm$51
kms$^{-1}$}. (See J. van de Sande et al. in preparation).

The dynamical mass is derived using
\begin{equation} M_{\rm dyn}=\frac{\beta(n)~ \sigma_{\star}^2~r_e }{G}
\label{eq:mdyn}
\end{equation} where $\beta(n)$ is an analytic expression as a
function of the S\'ersic index, as described by
\citet{cappellari2006}. Using $n=5.27$, we find $\beta=5.16$, and a
dynamical mass for NMBS-C7447 of 1.7$\pm0.5 \times $10$^{11}$ \msun.


\begin{figure*}[tbh] \epsscale{1.1} \plotone{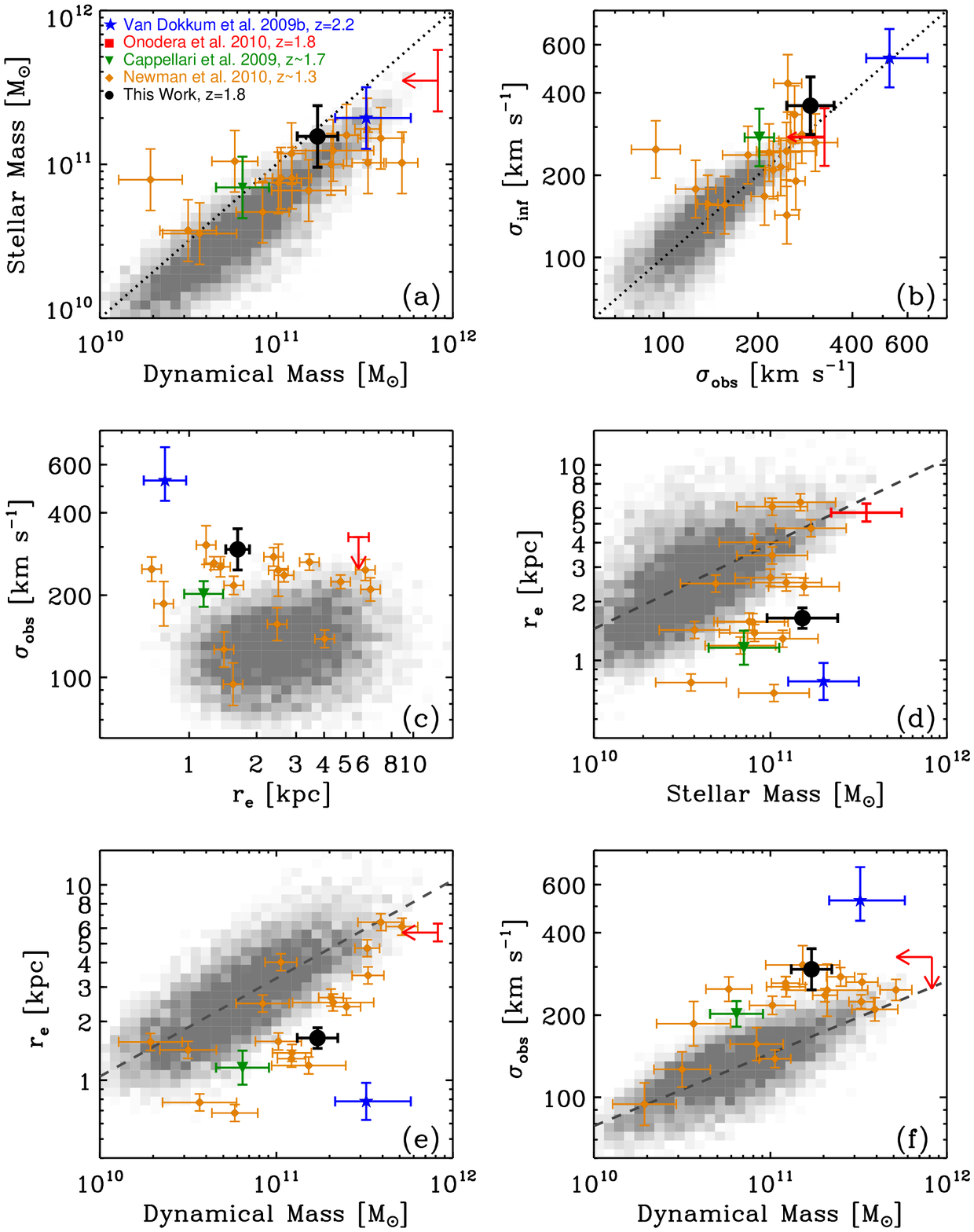}
\caption{ Comparison of NMBS-C7447 (black filled circle) with other
high-redshift studies (see legend in panel \textbf{a}) and quiescent
low-redshift galaxies in the SDSS (grey scale). \textbf{(a)}:
Dynamical vs. stellar mass. The dynamical and stellar masses are
consistent, but differ by a constant factor, which can be due to dark
matter and systematic effects on the stellar mass
estimates. \textbf{(b)} Measured vs. inferred velocity dispersions (as
inferred from the stellar mass and size). The observed dispersion
agrees well with inferred dispersion, which implies that the stellar
mass and size measurements are robust. \textbf{(c)}: $\sigma_{\rm obs}$ vs. $r_{\rm e}$. 
NMBS-C7447 is offset from the low-redshift
galaxies plane. \textbf{(d\&e)}: $r_{\rm e}$ vs. stellar and dynamical
mass. The dashed grey lines are the best-fit low-redshift relations
(Equation \ref{eq:mdre}). NMBS-C7447 is a factor of $\sim$2.5 smaller
than low-redshift galaxies at fixed mass. \textbf{(f)}: $\sigma_{\rm
obs}$ vs. dynamical mass. The velocity dispersion of NMBS-C7447 is a
factor $\sim$1.8 higher than similar-mass low-redshift galaxies.}
\label{fig:all_relations}
\end{figure*}

\begin{figure*}[tbh] \epsscale{1.15} \plotone{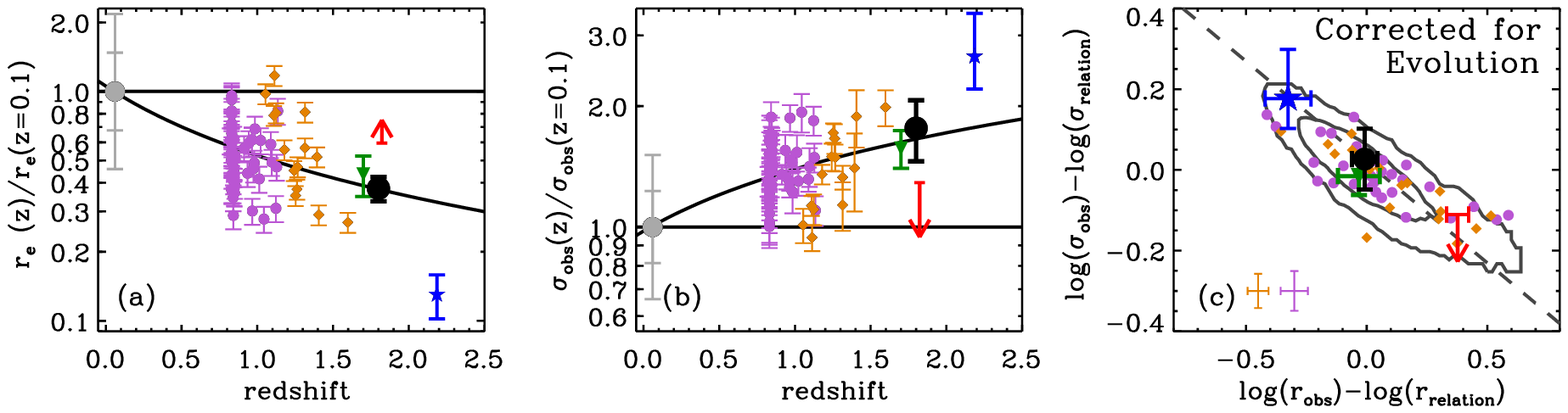}
\caption{Evolution in effective radius and velocity dispersion at
fixed dynamical mass, thus corrected for the $M_{\rm dyn}-r_e$ and
$M_{\rm dyn}-\sigma_{\rm obs}$ relations from Figure
\ref{fig:all_relations}e$\&$f.  We only select galaxies with
$M_{\rm dyn}>3\times10^{10}$\msun. Symbols are as described in Figure
\ref{fig:all_relations}, with the addition of data from
\citealt{vanderwel2008} (purple circles).  The grey filled circle at
$z\sim0.06$ shows the median from the SDSS, with the error indicating
the 1-, and 2-$\sigma$ scatter. The solid lines show a simple best
fit to the data of $(1+z)^{-0.98\pm0.09}$ for the evolution in
effective radius, and $(1+z)^{0.51\pm0.07}$ for the velocity
dispersion. \textbf{(c)} Scatter in the $M_{\star}-r_e$
relation versus the scatter in the $M_{\rm dyn}-\sigma_{\rm obs}$
relations, together with the 1- and
2-$\sigma$ contours of the SDSS galaxies, all corrected for evolution.
The discrepancy between the different measurements is expected
based on the intrinsic scatter in the low-redshift relations. }
\label{fig:redshift_delta}
\end{figure*}
%

\section{Evolution}
\label{subsec:implications}

In this section we compare our results to low- and high-redshift
measurements, and discuss the implications for the evolution of
quiescent galaxies. Figure \ref{fig:all_relations}
shows our results, together with other kinematical studies at $z>1$,
and galaxies from the SDSS at $0.05<z<0.07$ \citep{york2000}. The SDSS
structural parameters are from \citet{franx2008}, though we only
select non-starforming galaxies (i.e. specific SFR $ < 0.3/t_H$, see
\citealt{williams2009}). For all galaxies, velocity dispersions were
corrected as described in Section \ref{subsec:dispersions}, and
stellar masses were converted to a \citet{chabrier2003} IMF. All
dynamical masses were derived using Equation \ref{eq:mdyn}.

Many high-redshift studies rely on photometric stellar masses, which
suffer from large uncertainties \citep[e.g.,][]{conroy2009}. Here, we
test these stellar masses by comparing them to our dynamical
measurements (Figure \ref{fig:all_relations}a). The dynamical and
stellar mass for NMBS-C7447 are in agreement, and consistent with
the relation for low-redshift galaxies. Given this good agreement, we
should be able to predict the velocity dispersion from the size and
stellar mass measurements.

We assume a constant ratio of $M_{\rm dyn}/M_{\star}=1.68$, which is
the average ratio for the SDSS sample, to account for dark matter and systematic uncertainties in the stellar mass estimate. We show the results
in Figure \ref{fig:all_relations}b. The predicted velocity
dispersions of NMBS-C7447, and the other $z>1.5$ galaxies are
consistent with the observed velocity dispersions.  This illustrates
the robustness of our size and mass measurements.

In Figure \ref{fig:all_relations}c we show the velocity dispersion
versus the effective radius. Similar to what has been found for other
high-redshift studies, NMBS-C7447 has a clear offset from the
low-redshift galaxy population. Its velocity dispersion is higher
compared to $z\sim0.06$ galaxies with similar radii.  The mass size
relation is shown in Figure \ref{fig:all_relations} d$\&e$. The
effective radius of our galaxy is smaller compared to local galaxies
at similar masses, confirming other studies at high redshift. From
Figure \ref{fig:all_relations}f, where we show the dynamical mass
versus the observed velocity dispersion, we find that NMBS-C7447 has a
higher velocity dispersion than similar-mass SDSS galaxies, in agreement
with other studies of high-redshift compact galaxies.
 
We parametrize the mass-size relation by
(\citealt{shen2003},\citealt{vanderwel2008}):
\begin{equation} r_e=r_c \left( \frac{M_{\rm dyn}}{10^{11}M_{\odot}}
\right)^b.
 \label{eq:mdre}
\end{equation} Using a least-squares fit to the low-redshift galaxy
sample, we find $r_c=3.32$kpc, and $b=0.50$.  When comparing
NMBS-C7447 to local galaxies at fixed dynamical mass, we find that the
effective radius is a factor $\sim$2.5 smaller. We use a similar
approach for the velocity dispersion as a function of dynamical mass
(Figure \ref{fig:all_relations}f):
\begin{equation} \sigma_{\star}=\sigma_c \left( \frac{M_{\rm
dyn}}{10^{11}M_{\odot}} \right)^b,
\label{eq:mdsigma}
\end{equation}
with $\sigma_c=145$kms$^{-1}$, and $b=0.26$. NMBS-C7447 has a higher
velocity dispersion by a factor $\sim1.8$ compared to the low-redshift
relation.

Figure \ref{fig:redshift_delta} shows the evolution of the sizes and
velocity dispersions of the galaxies, normalized to a standard
dynamical mass using Equations \ref{eq:mdre} and \ref{eq:mdsigma}. We
add the sample by \citet{vanderwel2008} for a more complete
redshift coverage, with stellar masses derived from the FIRES \citep{forster2006} and FIREWORKS catalog \citep{wuyts2008}. We use a simple power law fit $r_e \propto
(1+z)^\alpha$ for galaxies with $M_{\rm dyn}>3\times10^{10}$\msun\ and find
$\alpha=-0.98\pm0.09$. This is in agreement with
\citet{vanderwel2008}, but slightly higher than
\citet{newman2010}. Our results imply a growth in size at fixed mass
by a factor of $\sim2.5$ from $z\sim1.8$ to the present day. When
assuming a similar power law for the velocity dispersion ($\sigma_{\star} \propto (1+z)^{0.51\pm0.07}$), we find a decrease
in velocity dispersion by a factor of $\sim1.5$ from $z\sim1.8$ to the
present day at fixed mass.

Figures \ref{fig:redshift_delta}a and b show that the scatter in the
relation in the normalized size and velocity dispersion is large at
fixed redshift. At $z>1.5$, three galaxies have been
observed with a range in normalized dispersions of a factor of
$\sim2.5$. This may lead to the conclusion that the measurements have
large unidentified errors and cannot be trusted yet.  On the other
hand, intrinsic scatter in the galaxy properties may cause this rather
large observed scatter. We can test this directly by using the
deviations of the galaxies in the mass-size relation.

If the scatter is due to variations in the intrinsic properties, we
expect that the deviations of the galaxies in the mass-size relation
correlate with the deviations of the galaxies in the mass-dispersion
relation. If the scatter is observational, there is no expected
correlation.  In Figure \ref{fig:redshift_delta}c, we compare the
deviation from the $M_{\star}-r_e$ relation to the deviations in the
$M_{\rm dyn}-\sigma_{\rm obs}$ relation.  
The deviation of the $M_{\star}-r_e$ and $M_{\rm dyn}-\sigma_{\rm obs}$
relations were derived using the evolution of these relations at fixed mass as
shown in Figures  \ref{fig:redshift_delta}a and b.

We can predict, using the virial
theorem, how the points would lie if the scatter is intrinsic, i.e.,
due to variations in the galaxy structure. This line is shown in the
Figure \ref{fig:redshift_delta}c, and we see that the galaxies lie
very close to this line.  In addition, we show the area which is
covered by the SDSS galaxies in the same diagram (1- and 2-$\sigma$
contours). Almost all data points lie within these contours. Hence we
conclude that the scatter is mostly {\sl intrinsic}, and not
observational. A direct measure of the average offset of the sizes and
dispersions can be obtained by increasing the number of observed
galaxies to about 30, which would reduce the error by a factor of
$\sim3$. Alternatively, the average mass-size relation can be used to
determine the average offset at $z=1.5-2$.  Thus we conclude that the
difference between our results and those by \citet{vandokkum2009b} and
\citet{onodera2010} are due to intrinsic scatter in galaxy properties.

\vspace{10pt}

\section{CONCLUSIONS} In this paper we have presented the first
high-S/N, high-resolution, spectrum of a compact massive quiescent
galaxy at $z=1.80$ observed with X-Shooter. Using this spectrum we
have determined the stellar mass and velocity dispersion: $M_{\star}
\sim1.5\times10^{11}$\msun, $\sigma_{\rm obs}=294\pm 51$km$~
$s$^{-1}$. From HST-WFC3 imaging we find that $r_e=1.64\pm0.15$kpc. The
stellar mass and dynamical mass agree well ($M_{\rm
dyn}=1.7\pm0.5\times10^{11}$\msun), and are consistent with the local
SDSS relation. Our results suggest that stellar masses at high
redshift are robust, and thus supports the claim that massive,
quiescent galaxies with high stellar mass densities at $z\sim2$ exist.

When comparing this galaxy to low-redshift early-type galaxies, we
find that it is structurally different. At fixed dynamical mass,
NMBS-C7447 is smaller by a factor $\sim2.5$, and has a higher velocity
dispersion by a factor of $\sim1.8$.

Despite the high accuracy of our derived stellar parameters, our study
is still limited to a single high-redshift galaxy, and it brings the
total number of stellar kinematic measurements for individual galaxies at $z>1.5$ to three.
We have shown that the differences between the three measurements can
be explained by the scatter in the mass-size relation. A
larger sample of compact massive quiescent galaxies at high redshift
is needed to accurately measure the structural evolution of these
galaxies with cosmic time.


\acknowledgments 
We thank the anonymous referee for his comments, Johan Fynbo and Joanna Holt on the reduction of the UVB and VIS X-Shooter data, and the NMBS team for their
contribution. This research was supported by grants from the
Netherlands Foundation for Research (NWO), the Leids Kerkhoven-Bosscha
Fonds. Support for program HST-GO-12167.1 was provided by NASA through
a grant from the Space Telescope Science Institute.


\clearpage

\end{document}